\documentclass[aps,twocolumn,groupedaddress]{revtex4}

\usepackage{amsmath}
\usepackage{epsfig}
\usepackage{amssymb}
\usepackage{nicefrac}

\newcommand{\nn}{\nonumber}

\def\a{\alpha}

\def\e{\epsilon}

\def\h{\eta}

\def\CM{{M^4}}

\def\IC{\relax\hbox{$\inbar\kern-.3em{\rm C}$}}

\def\bM{{\bf M}}

\def\IC{{\bf C}}

\def\bea{\begin{eqnarray}}
\def\eea{\end{eqnarray}}
\def\be{\begin{eqnarray}}
\def\ee{\end{eqnarray}}
\def\ba{\begin{align}}
\def\ea{\end{align}}
\def\bse{\begin{subequations}}
\def\ese{\end{subequations}}
\def\1F1{{}_1\!F_1}
\def\2F0{{}_2\!F_0}

\def\ni{\noindent}

\def\bx{\bar{x}}
\def\by{\bar{y}}

\def\nn{\nonumber}

\def\a{\alpha}

\def\h3{$\textrm{H}_3^+$}

\def\IC{{\mathbb C}}

\def\lbldef#1#2{\expandafter\gdef\csname #1\endcsname {#2}}

\def\href#1#2{#2}

\newcommand{\beq}{\begin{equation}}
\newcommand{\eeq}{\end{equation}}
\newcommand{\ber}{\begin{eqnarray}}
\newcommand{\eer}{\end{eqnarray}}

\def\mo{{\mathbb{O}}}
\def\mv{{\mathbb{V}}}
\def\bM{{\bar{M}}}
\def\bJ{{\bar{J}}}

\def\e{{\epsilon}}
\def\eb{{\bar{\epsilon}}}
\def\a{{\alpha}}
\def\ba{{\bar{\alpha}}}
\def\+{\!+\!}
\def\-{\!-\!}
\def\={\!=\!}

\def\({\left(}
\def\){\right)}
\def\[{\left[}
\def\]{\right]}
\def\<{\langle}
\def\>{\rangle}

\begin{document}

\hbox{ \small YITP-SB-07-15, BRX-TH-587.}
\vskip .5 cm
\title{Exact $N=4$ correlators of  AdS$_3/$CFT$_2$}


\author{Ari Pakman\footnote{ari.pakman@stonybrook.edu}}
\affiliation{ C.N.Yang Institute for Theoretical Physics,
Stony Brook University, Stony Brook, NY 11794-3840, USA }

\author{Amit Sever\footnote{asever@brandeis.edu}}
\affiliation{Brandeis Theory Group, Martin Fisher School of Physics, Brandeis University, Waltham, MA, 02454-9110, USA}


\begin{abstract}
\ni
We extend to chiral $N=4$ operators the holographic
agreement recently found between correlators of the symmetric orbifold  of $M^4$ at large $N$
and type IIB strings
propagating in $AdS_3 \times S^3 \times M^4$, where $M^4=T^4$ or $K3$.
We also present expressions for some bulk correlators not yet computed in the boundary.
\end{abstract}


\maketitle

\section{Introduction}
One of the simplest realizations of the
$AdS_{n+1}/CFT_n$ duality \cite{Maldacena:1997re, Gubser:1998bc, Witten:1998qj,
Aharony:1999ti} is the duality between type IIB string theory
in  $AdS_3 \times S^3 \times \CM$,
where $\CM$ is either a torus $T^4$
or a $K3$ surface, and a two-dimensional $N=4$ superconformal field theory
in the moduli space of
the symmetric product of $\CM$.
This duality can be derived by considering the near horizon of
a system of $Q_1$ D1 branes and $Q_5$ D5 branes wrapping $M^4$.

The bulk and the boundary theories have equivalent  moduli spaces \cite{Dijkgraaf:1998gf,Larsen:1999uk},
and on both sides of the duality there are special points
where the theory has a solvable description.
In the bulk, the special point  corresponds  to a supergravity frame without
RR flux~\cite{Maldacena:1998bw}, where the string worldsheet  is
described, for Euclidean $AdS_3$,  by  \h3\ and $SU(2)$ WZW models at level $k=Q_5$.
The second special point corresponds
in the boundary to the symmetric orbifold of $N=Q_1Q_5$ copies of $M^4$.

Recently, progress was made in checking the duality of the two theories
at the dynamical level \cite{Gaberdiel:2007vu,Dabholkar:2007ey} by comparing
correlators at these solvable points.
It was shown there that the large $N$ limit of certain three-point functions of chiral fields
computed earlier in the symmetric product CFT
agree precisely with string theory three-point functions computed in the sphere.
This verification of the $AdS_3/CFT_2$ duality  is surprising  because the
computations are carried out at different points in the moduli space,
thus suggesting a non-renormalization theorem.

In \cite{Gaberdiel:2007vu} it was shown that computations in the bulk
reproduce one of the correlators
of chiral $SU(2)$ multiplets computed in the boundary in \cite{Lunin:2000yv,Lunin:2001pw}.
In \cite{Dabholkar:2007ey} it was shown that, for $M^4=T^4$, the fusion rules and the structure constants
of the complete $N=2$ chiral ring in the bulk are in precise agreement with the boundary
results of~\cite{Jevicki:1998bm}.

In this note we show that a simple computation allows to
extend the comparison  
to those cases 
not considered in \cite{Gaberdiel:2007vu,Dabholkar:2007ey}.
We will show that the agreement of correlators for chiral $N=4$ multiplets
holds for all the boundary correlators computed in \cite{Lunin:2001pw}.
In addition,  we will give expressions for three-point functions
in the bulk for $M^4=T^4$ which were not yet computed in the boundary.

\section{Bulk-Boundary Agreement  \label{bulk}}
Chiral $SU(2)$ multiplets in $AdS_3 \times S^3 \times T^4$
are operators satisfying $H=J$, where $J$ is the $SU(2)$ spin and $H$ is the $SL(2,R)$ spin,
which is interpreted as the conformal dimension in the dual theory.
Physical string operators of this kind
are given, in the  holomorphic sector, by three families \cite{Kutasov:1998zh}, as shown in the table below.
\begin{center}
\begin{tabular}{|l|c|l|c|}
\hline
Field        & $H= J  $ & Range of $H$ & Sector
\\ \hline \hline
$\mo_{h}^{0}$ &  $h-1=j$  & $0,\nicefrac12 \ldots \frac{k-2}{2}$ & NS\\  \hline
$\mo_{h}^{a}$ &  $h-1/2=j+1/2$ & $\nicefrac12,1 \ldots \frac{k-1}{2}$ & R \\ \hline
$\mo_{h}^{2}$ &  $h=j+1$  & $1,\nicefrac32 \ldots \nicefrac{k}{2}$ & NS  \\ \hline
\end{tabular}
\end{center}
Here  $a=1,2$ correspond to the two holomorphic one-forms in $T^4$. The numbers $h,j$ are the spins of the operators under
the bosonic $SL(2,R)_{k+2}$ and $SU(2)_{k-2}$
which appear in the decomposition of the supersymmetric WZW models
into bosonic WZW models and free fermions (see~\cite{Dabholkar:2007ey} for details).
The number $h$ takes the $k\-1$ values $h\=1, \nicefrac32 \ldots \nicefrac{k}{2}$.
For a given $h$, each operator has also an anti-holomorphic label,
so the full operators are, e.g.,~$\mo^{(0,2)}_h$, etc.

The same  families of operators appear in the boundary theory \cite{Jevicki:1998bm},
but the range of $H$ there  is larger. It is expected that
additional operators in the bulk come from including spectrally flowed sectors
of~$SL(2,R)$~\cite{Maldacena:2000hw,Hikida:2000ry,Argurio:2000tb},
which we will not consider here.

The operators depend on the variables $x, \bx$, which are interpreted as the local variables of the
boundary theory, and on $y,\by$, which are isospin $SU(2)$ variables \cite{Zamolodchikov:1986bd}.
They are normalized as
\be
\langle
\mo_h^{(\a,\ba)} \mathbb{O}_h^{(\a,\ba)} \rangle
= \frac{(y_1-y_2)^{2J} (\bar{y}_1 - \bar{y}_2)^{2\bar{J}}}{(x_1-x_2)^{2H} (\bx_1-\bx_2)^{2\bar{H}}}\,,
\ee
and
can be expanded  into  modes with
definite $J^3_0, \bar{J}^3_0$ eigenvalues,
\be
 \mo_h^{(\a,\ba)}(y,\by)&=& \!
 \sum_{M=-J}^{J} \sum_{\bM=-\bJ}^{\bJ}
\! \(c_{M}^{J} c_{\bM}^{\bJ}\)^{\nicefrac{1}{2}}
\nn
\\
 &\times  &y^{\-M+J} \bar{y}^{-\!\bM+\bJ} \mv_{h,M,\bM}^{(\a,\ba)} \,,
\label{vm}
\ee
where
\be
c_{M}^{J} = \(\begin{array}{c} 2J \\ M+J \end{array}\) = \frac{(2J)!}{(J+M)!(J-M)!} \,.
\ee
The modes $\mv_{h,M,\bM}$ are normalized as
\be
\<\mv_{h,M,\bM}^{(\a,\ba)} \mv_{h,-M,-\bM}^{(\a, \ba)} \rangle = (-1)^{J + \bJ \- M \- \bM} \,,
\ee
where we have taken $x_1\=\bx_1\=1, x_2\=\bx_2\=0$.

The string theory three-point functions  for chiral operators
were shown in~\cite{Dabholkar:2007ey} to be
\be
&
\langle
\mo_{h_1}^{(\!\a_1,\ba_1 \!)}  \mo_{h_2}^{(\!\a_2,\ba_2\!)}
\mo_{h_3}^{(\!\a_3,\ba_3\!)}   \rangle  =
\frac{N^{-\frac12} f(h_i;\a_i) f(h_i;\ba_i)}{\sqrt{(2h_1\-1)(2h_2\-1)(2h_3\-1)}}
\label{3pf}
\\
 \nn
 & \times \,\,
y_{12}^{J_1+J_2-J_3}
y_{23}^{J_2+J_3-J_1}
y_{31}^{J_3+J_1-J_2}
 \\
 \nn
& \times \,\,
\by_{12}^{\bar{J}_1+\bar{J}_2-\bar{J}_3}
\by_{23}^{\bar{J}_2+\bar{J}_3-\bar{J}_1}
\by_{31}^{\bar{J}_3+\bar{J}_1-\bar{J}_2} \,,
\ee
where $y_{12}=y_1-y_2$, etc.,
the operators are at
$x\=0, 1,\infty$, and the functions
$f(h_i;\a_i)=f(h_i;\a_1,\a_2,\a_3)$ are given by
\be
f(h_i;0,0,0) &=& -h_1-h_2-h_3+2
\nn
\\
f(h_i;0,0,2) &=& -h_1 - h_2 + h_3+1
\nn
\\
f(h_i;0,2,2) &=&  -h_1 + h_2+h_3
\label{fa}
\\
f(h_i;2,2,2) &=& h_1+h_2+h_3-1
\nn
\\
f(h_i;0,a,b) &=&  f(h_i;2,a,b)
\nn
\\
\nn
&=&  \sqrt{( 2h_2 \-1 )(2h_3 \-1 )}\xi^{ab} \,,
\ee
with  $\xi^{12}\=\xi^{21}\=1, \xi^{11}\=\xi^{22}\=0$.
Note that all the dependence on the type of operator $\a_i, \ba_i$
is encoded in the functions $f(h_i;\a_i)$ and
is completely factorized in (\ref{3pf}) between holomorphic and anti-holomorphic sectors.

In \cite{Dabholkar:2007ey} only  $N=2$ chiral states were considered, so the relation $J_3 = J_2 + J_1$ was imposed
and only the $M_{1,2}=J_{1,2}$, $M_3=-J_3$ members of the $SU(2)$ multiplet were retained
by taking the limits $y_{1,2} \rightarrow 0, y_3 \rightarrow \infty$.
Here we will keep both $J_i$'s and $y_i$'s arbitrary, the only restriction coming from the $SU(2)$ fusion rules
applied to the $j_i$'s
and U(1) R-charge conservation $M_1\+M_2\+M_3\=0$
(and similarly for the $\bM_i$'s).
This case was considered in~\cite{Gaberdiel:2007vu} for correlators with~$\a_i\=\ba_i\=0$,
and~$M\=\bM$. In this note, we consider  arbitrary $\a_i,\ba_i=0,2$ and $M,\bM$.
Our results will thus be valid for both $M^4=T^4$ and $M^4=K3$, since
only for operators with $\a,\ba=a$ these two cases differ.
Correlators involving $\a,\ba=a$ $N=4$ chiral
primaries with $J_3 < J_1+J_2$ were not computed yet in the boundary conformal field theory,
so for these cases we will present the predictions from the bulk for $M^4=T^4$.

Let us express the operators in terms of
\be n = 2h-1\,,
\ee
where, in the symmetric orbifold, $n$ is the length of the permutation
cycle in the corresponding operator.
Let us label also the two types of operators by $\e=-1$ for $\a=0$ and $\e=+1$
for $\a=2$. The spins are given now by
\be J_i = \frac{n_i+\e_i}{2}
\qquad \bJ_i = \frac{n_i+\eb_i}{2} \,.
\label{jjb}
\ee
Remarkably, all the correlators with $\a=0,2$,
which were computed in \cite{Dabholkar:2007ey}
separately for each case, can be
expressed in terms of $n_i,\e_i$ in a symmetric form as
\begin{eqnarray}
&\langle \mo_{n_1}^{(\e_1,\eb_1)}  \mo_{n_2}^{(\e_2,\eb_2)}
\mo_{n_3}^{(\e_3,\eb_3)}   \rangle =
\label{sym}
\\
\nn
&\frac{1}{\sqrt{N}}\frac{(\e_1 n_1 + \e_2n_2 + \e_3n_3 +1 ) (\eb_1 n_1 + \eb_2n_2 + \eb_3n_3 +1 ) }
{4(n_1 n_2 n_3)^{\nicefrac12}}
\\
 \nn
& \times \,\,
y_{12}^{J_1+J_2-J_3}
y_{23}^{J_2+J_3-J_1}
y_{31}^{J_3+J_1-J_2}
 \\
 \nn
& \times \,\,
\by_{12}^{\bar{J}_1+\bar{J}_2-\bar{J}_3}
\by_{23}^{\bar{J}_2+\bar{J}_3-\bar{J}_1}
\by_{31}^{\bar{J}_3+\bar{J}_1-\bar{J}_2} \,.
 \label{PD}
\end{eqnarray}
To compare with the results of \cite{Lunin:2001pw} we should recast
this expression  in the $M,\bM$ basis.
Expanding~(\ref{sym}) using~(\ref{vm}), it is easy to read out the term
\be
& \<\mv_{n_1,-\!J_1,-\!\bJ_1}^{(\e_1,\eb_1)}\mv_{n_2,J_2,\bJ_2}^{(\e_2,\eb_2)}\mv_{n_3,J_1\!-\!J_2,\bJ_1\!-\!\bJ_2}^{(\e_3,\eb_3)}\> =
\\
\nn
&
\frac{N^{-\frac12}}{\(c^{J_3}_{J_1\!-\!J_2} c^{\bJ_3}_{\bJ_1\!-\!\bJ_2}\)^{\nicefrac12}}
\frac{(\e_1 n_1 + \e_2n_2 + \e_3n_3 +1 ) (\eb_1 n_1 + \eb_2n_2 + \eb_3n_3 +1 ) }
{4(n_1 n_2 n_3)^{\nicefrac12}} \,,
\ee
where we have used $\(\-1\)^{2(J_3+\bJ_3)} \=1$, as follows from (\ref{jjb}).
The general correlator in the $M, \bM$ basis follows from the  Wigner-Eckart theorem and is given by
\be
&\<\mv_{n_1,M_1,\bM_1}^{(\e_1,\eb_1)}\mv_{n_2,M_2,\bM_2}^{(\e_2,\eb_2)}\mv_{n_3,M_3,\bM_3}^{(\e_3,\eb_3)}\>
=
\\ \nn
&\<\mv_{n_1,\-J_1,\-\bJ_1}^{(\e_1,\eb_1)}\! \mv_{n_2,J_2,\bJ_2}^{(\e_2,\eb_2)}\mv_{n_3,J_1\-J_2,\bJ_1\-\bJ_2}^{(\e_3,\eb_3)}\>
\\ \nn
& \times
\frac{ d^{J_1,J_2,J_3}_{M_1,M_2,M_3} d^{\bJ_1,\bJ_2,\bJ_3}_{\bM_1,\bM_2,\bM_3}}
{ d^{J_1,J_2,J_3}_{\-J_1,J_2,J_1\-J_2} d^{\bJ_1,\bJ_2,\bJ_3}_{\-\bJ_1,\bJ_2,\bJ_1\-\bJ_2} } \,,
\ee
where
\be
d^{J_1,J_2,J_3}_{M_1,M_2,M_3} = \(\begin{array}{ccc} J_1&J_2&J_3 \\M_1&M_2&M_3 \end{array}\)
\ee

\ni
are the $SU(2)$ 3j symbols.
Using now
\be
 d^{J_1,J_2,J_3}_{\-J_1,J_2,J_1\-J_2} =
 \[ \frac{(2J_1)!(2J_2)!}{(J_2 \+ J_2 \- J_3 )! (J_1\+J_2\+J_3\+1)! } \]^{\nicefrac12}
\ee
we get
\be
&\<\mv_{n_1,M_1,\bM_1}^{(\e_1,\eb_1)}\mv_{n_2,M_2,\bM_2}^{(\e_2,\eb_2)}\mv_{n_3,M_3,\bM_3}^{(\e_3,\eb_3)}\>
=
\label{final}
\\
\nn
& L(J_i,M_i) L(\bJ_i,\bM_i)
\\
\nn
\times & \frac{1}{\sqrt{N}}\frac{(\e_1 n_1 + \e_2n_2 + \e_3n_3 +1 ) (\eb_1 n_1 + \eb_2n_2 + \eb_3n_3 +1 ) }
{4(n_1 n_2 n_3)^{\nicefrac12}}
\ee
where
\be
& L(J_i,M_i) =
d^{J_1,J_2,J_3}_{M_1,M_2,M_3}
\\
\nn
\times &
\[ \frac{(J_1+J_2-J_3)! (J_2+J_3-J_1)! (J_3+J_1-J_2)! (J_1+J_2+J_3+1)!}
{(2J_1)!(2J_2)!(2J_3)!}\]^{\nicefrac12} \,.
\ee
Eq.(\ref{final}), which is the main result of this note,
coincides precisely with eq.(6.47) of~\cite{Lunin:2001pw},
with the identifications $n_1 \= n, n_2 \= m, n_3 \= q, \e_1\=1_n, \e_2\=1_m, \e_3\=1_q$.

Correlators involving operators with $\a, \ba \= a$ are expressed
similarly in the $M, \bM$ basis using~(\ref{fa}).
There are essentially three classes of such correlators, given by
\be
&
\hspace{-3cm}
\<\mv_{n_1,M_1,\bM_1}^{(a,\eb_1)}\mv_{n_2,M_2,\bM_2}^{(b,\eb_2)}\mv_{n_3,M_3,\bM_3}^{(\e_3,\eb_3)}\>
= \\
\nn
 &
\hspace{1cm}
L(J_i,M_i) L(\bJ_i,\bM_i) \times \frac{1}{\sqrt{N}}\frac{\xi^{ab}(\eb_1 n_1 + \eb_2n_2 + \eb_3n_3 +1 ) } {2( n_3)^{\nicefrac12}}
\,,
\\
& \hspace{-3cm}
\<\mv_{n_1,M_1,\bM_1}^{(a,\bar{a})}\mv_{n_2,M_2,\bM_2}^{(b,\bar{b})}\mv_{n_3,M_3,\bM_3}^{(\e_3,\eb_3)}\>
=
\\
\nn
 &
 \hspace{1cm}
L(J_i,M_i) L(\bJ_i,\bM_i) \times \frac{1}{\sqrt{N}} \xi^{ab}\xi^{\bar{a}\bar{b}}
\(\frac{n_1 n_2}{n_3} \)^{\nicefrac12}
\,,
\\
& \hspace{-3cm}
\<\mv_{n_1,M_1,\bM_1}^{(a,\eb_1)}\mv_{n_2,M_2,\bM_2}^{(b,\bar{b})}\mv_{n_3,M_3,\bM_3}^{(\e_3,\bar{a})}\>
=
\\
\nn
& \hspace{1cm}
L(J_i,M_i) L(\bJ_i,\bM_i) \times \frac{1}{\sqrt{N}} \xi^{ab}\xi^{\bar{a}\bar{b}}
(n_2)^{\nicefrac12}
\,.
\ee
It would be interesting to
extend the computations of~\cite{Lunin:2001pw}
in order to verify the holographic agreement for these correlators.

\section{Discussion  \label{discussion}}
The bulk-boundary agreement
found  is surprising because
the computations are done at two largely separated points in the moduli space,
suggesting a non-renormalization theorem which should be investigated.
Since the agreement found here is valid at
large $N$, the question arises
whether such non-renormalization theorem would hold also at finite $N$,
and if so, how the finite $N$ corrections should be obtained in the
bulk \footnote{Note that in the boundary, these finite $N$ corrections come
from a covering space with the topology of a sphere.}.
Among other interesting open questions,
in \cite{Dabholkar:2007ey} it was pointed
out that for chiral operators in the boundary there are several ways
of combining the fermions which multiply  the twist fields.
It would be interesting to understand
what these options correspond to in the bulk.

\section*{Acknowledgements}
\ni
We thank Oleg Lunin  for conversations that led to this
work and David Kutasov, Albion Lawrence, William Linch and John McGreevy for useful discussions.
AP thanks the University of Chicago
and AS thanks the MIT Center for Theoretical Physics for
their generous hospitality.
The work of AP is supported by the Simons Foundation.
The work of AS is supported in part by the National
Science Foundation under Grant PHY-0331516,
by DOE Grant No. DE-FG02-92ER40706
and by an Outstanding Junior Investigator award.

\bibliography{h3bib}

\end{document}